\documentclass{article}
\usepackage{arxiv}

\usepackage[utf8]{inputenc} 
\usepackage[T1]{fontenc}    
\usepackage{hyperref}       
\usepackage{url}            
\usepackage{booktabs}       
\usepackage{amsfonts}       
\usepackage{nicefrac}       
\usepackage{microtype}      
\usepackage{lipsum}		
\usepackage{graphicx}
\usepackage{doi}

\usepackage{graphicx}
\usepackage{amssymb, amsmath}
\usepackage{color}
\usepackage{url}
\usepackage{comment}
\usepackage{afterpage}
\usepackage{here}
\newcommand{\reffig}[1]{Figure~\ref{#1}}

\newcommand{\reftab}[1]{Table~\ref{#1}}
\newcommand{\refeq}[1]{Eq. \eqref{#1}}
\newcommand{\bi}[1]{\ensuremath{\boldsymbol{#1}}}

\title{Designing ship hull forms using generative adversarial networks}

\author{ \href{https://orcid.org/0000-0002-1955-069X}{\includegraphics[scale=0.06]{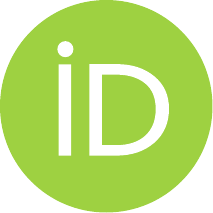}\hspace{1mm}Kazuo Yonekura}\\
	The University of Tokyo,\\
	7-3-1 Hongo, Bunkyo-ku, Tokyo, Japan 113-8656 \\
	\texttt{yonekura@struct.t.u-tokyo.ac.jp} \\
	\And
		Kotaro Omori \\
		The University of Tokyo,\\
		7-3-1 Hongo, Bunkyo-ku, Tokyo, Japan 113-8656 \\
		\And
		Xinran Qi \\
	The University of Tokyo,\\
	7-3-1 Hongo, Bunkyo-ku, Tokyo, Japan 113-8656 \\
	\And
		Katsuyuki Suzuki \\
	The University of Tokyo,\\
	7-3-1 Hongo, Bunkyo-ku, Tokyo, Japan 113-8656 \\
}

\renewcommand{\shorttitle}{\textit{arXiv} Template}

\hypersetup{
pdftitle={},
pdfsubject={},
pdfauthor={},
pdfkeywords={},
}

\begin{document}
\maketitle

\begin{abstract}
We proposed a GAN-based method to generate a ship hull form. Unlike mathematical hull forms that require geometrical parameters to generate ship hull forms, the proposed method requires desirable ship performance parameters, i.e., the drag coefficient and tonnage. 
The requirements of ship owners are generally focused on the ship performance and not the geometry itself. Hence, the proposed model is useful for obtaining the ship hull form based on an owner's requirements. 
The GAN model was trained using a ship hull form dataset generated using the generalized Wigley hull form. The proposed method was evaluated through numerical experiments and successfully generated ship data with small errors.
\end{abstract}


	\section{Introduction}
\label{}
A mathematical or parametric hull form was developed to design or study ship hull forms \cite{Matsui22,jmse10050686,Wu01,Zhang08}. 	It has also been used for optimization \cite{Feng21,Tahara06,Feng21,Han12}.
In general, geometric parameters are specified in a mathematical hull form, and the ship hull form is defined based on these parameters. 
The geometric parameters include the block coefficient, midship area coefficient, and coordinates of B-spline curves. 	
The modified Wigley hull form is a well-known and widely used mathematical hull form in experimental and numerical studies \cite{Matsui22}. 
The mathematical hull form requires design parameters to define the hull form. 
For example, the generalized Wigley hull form requires the hull length, width, draft, block coefficient, prismatic coefficient, midship section coefficient, and water plane area coefficient. 
The mathematical hull form requires geometric parameters to define the ship hull. 
However, the requirements of ship owners, such as speed, tonnage, and fuel consumption rate, are often simpler, and the aforementioned geometric parameters are not directly linked to owners’ requirements. 
It is useful if the ship hull form is generated based on these simple requirements. 
In this study, we proposed a machine-learning-based method to generate hull forms by inputting the design speed, tonnage, and drag coefficient. 

Machine learning methods have recently been utilized to generate shapes \cite{Regenwetter22,Chaudhari23,Shu19}. 
A generative adversarial network (GAN) is a method used to generate, for example, laminar \cite{Achour20,Yonekura22b} and supercritical airfoils \cite{WANG202262}. 
A variational autoencoder (VAE) is another machine-learning method used for similar tasks \cite{Yonekura21a,Yonekura22c}. 
When generating mechanical shapes, one can specify the desired performance and generate shapes that meet the requirements using a conditional GAN or a conditional VAE. 
For example, in an airfoil generation task, the lift coefficient at a certain angle of attack is specified, and the airfoil is generated \cite{Yonekura22b,Yonekura22c}. 
Both 2D and 3D \cite{Shu19} shapes can be generated using the generative models. 
\cite{Khan23} proposed the ShipHullGAN that generates ship hull data; however, it does not consider performance parameters. Thus, the model only generated data, but the ship performance was unknown. 

By utilizing the GAN model for the ship hull form, a new ship hull form design method was proposed in this study. 
The proposed method uses a conditional Wasserstein GAN with a gradient penalty (cWGAN-gp), which is used in airfoil generation tasks \cite{Yonekura22b}. 
The generalized Wigley ship hull form was employed to generate the training data, which were then used to train the cWGAN-gp model. 
Once trained, the generator of the cWGAN-gp outputs new ship hull data by specifying the requirements. 
In the numerical examples, the drag coefficient and tonnage were employed as requirements. 
The trained GAN model can be used in the same manner as a mathematical hull form. 
Both models output the ship hull form. The mathematical hull form requires geometric parameters as inputs, whereas the GAN model utilizes performance requirements as inputs.
In addition, it has been reported that using various data as training data increases the variety of output data \cite{Yonekura22c}. 

The remainder of this paper is organized as follows. 
GAN models are introduced in Section 2. The ship hull form generation method is described in Section 3. We also introduce here the generalized Wigley hull form dataset and the computation method for the drag coefficient. 
Numerical examples are presented in Section 4, and the study is concluded in Section 5.

\section{Generative adversarial network}
GAN \cite{goodfellow2017nips} and conditional GAN (cGAN) \cite{mirza14} are composed of generator $G$ and discriminator $D$ networks. 
The generator network inputs a random noise vector $\bi{z}$, which is also called a latent vector, and a label vector $\bi{c}$. 
The generator network outputs fake data that mimic the training data, which are also called true data. 
The discriminator network distinguishes between true and fake data. 
The optimization problem of GAN is written as $\min_{G} \max_{D} V(G, D)$, where the loss function $V(G,D)$ is formulated as
\begin{align*}
	& V(D, G) = \mathbb{E}_{\bi{x} \sim p_r(\bi{x})} [ \log{D(\bi{x})} ]
	+ \mathbb{E}_{\bi{z} \sim p_z(\bi{z})} [\log{ (1-D(G(\bi{z})) ) } ] ,
\end{align*}
where $\bi{x}$ and $\bi{z}$ represent the training data and latent vector, respectively, and $p_r(\bi{x})$ and $p_z(\bi{z})$ represent their corresponding probability distributions. 
$G$ and $D$ are treated as functions, where the output of $G$ is the data. 
The output of $D$ is a scholar in $[0,1]$, where $0$ and $1$ correspond to fake and true data, respectively.
On one hand, function $V$ indicates a larger value if the discriminator succeeds in determining whether the input data are true or fake. 
On the other hand, $V$ indicates a smaller value if the generator succeeds in cheating on the discriminator.
Therefore, GAN is referred to as adversarial training. 

It has been reported that training a GAN is unstable owing to mode collapse \cite{goodfellow2017nips} and gradient dissipation \cite{Arjovsky17a}.
To overcome these problems, the Wasserstein GAN (WGAN) was proposed \cite{Arjovsky17b}, and the WGAN with a gradient penalty (WGAN-gp) \cite{WGANgp} further improved the stability.
In the WGAN, the Earth mover's distance (EM distance or Wasserstein distance) is employed to measure the difference between the two probability distributions of real and fake data. 
The EM distances of the two probability distributions $p_r$ and $p_g$ are formulated as
\begin{align*}
	W \left( p_r, p_g \right) = \sup_{ \| f \|_{\mathcal L} \leq 1 } 
	\mathbb{E}_{\bi{x} \sim p_r} [ f(\bi{x}) ] - \mathbb{E}_{\bi{x} \sim p_g} [ f(\bi{x}) ], 
\end{align*}	
In the first term on the right-hand side, function $f$, which denotes a neural network parameter of the discriminator network, must be a $1$-Lipshitz function. 
The loss function in the WGAN-gp is formulated as 
\begin{align*}
	&\mathcal{L}_{WGAN \mathchar` gp} = \mathbb{E}_{\tilde{x} \sim P_g} \left[ D \left( \tilde{x} \right) \right] 
	- \mathbb{E}_{ x \sim P_r} \left[ D \left( x \right) \right]
	+ \lambda \mathcal{L}_{gp} , \\
	&\mathcal{L}_{gp} = \mathbb{E}_{ \hat{x} \sim P_{\hat{x}} } \left[ \left( \left \| \nabla_{\hat{x}} D \left( \hat{x} \right) \right \|_2 -1 \right)^2 \right]. 
\end{align*}
The term $\mathcal{L}_{gp}$ is called the gradient penalty and is employed to ensure $1$-Lipshitzness.

\section{Ship hull-form design}
\subsection{Hull-form design task}
We focused on the preliminary design of ship hulls. 
The aim was to draw a ship hull form that satisfied the requirements specified in advance.
In the present study, we assume that the ship owner generally requires a specific displacement tonnage and ship speed and desires a better fuel consumption rate. 
Therefore, the design task involved drawing a line plan, in which the drag coefficient $C_d$, design speed $U$, and displacement tonnage $W$ were specified. 

The drag coefficient $C_d$ is the weighted sum of the frictional drag coefficient $C_{df}$ and wave drag coefficient $C_{dw}$. 
\begin{align}
	C_d = (1+K) C_{df} + C_{dw}, \label{eq.cd}
\end{align}
where $K$ denotes the form factor. 
\cite{Prohaska66} proposed a method to estimate $K$ as follows: 
\begin{align*}
	K = 0.11 + 0.128 \frac{B}{d} - 0.0157 \left( \frac{B}{d} \right) ^2
	- 3.1 \frac{C_b B}{L} + 28.8 \left( \frac{C_b B}{L} \right) ^2. 
\end{align*}
The frictional drag coefficient $C_{df}$ is given by 
\begin{align*}
	C_{df} = 1.328 {R_n}^{-\frac{1}{2}}. 
\end{align*}
The wave drag coefficient $C_{dw}$ is calculated as follows:
\begin{align*}
	C_{dw} = \frac{8}{\pi {F_n}^4}  \int_0 ^{\frac{\pi}{2} } \left[ P \left( \theta \right)^2 + Q \left( \theta \right)^2 \right] \sec ^3 \theta  ~{\rm d}\theta, 
\end{align*}
where $F_n$ is the Froude number, and $P$ and $Q$ are the amplitude functions. 
The Froude number is defined as 
$ F_n = \frac{U}{ \sqrt{gL}}$ using gravitational acceleration $g$. 
$P$ and $Q$ are defined as
\begin{align*}
	&	P(\theta) = \iint_{S_c} \frac{\partial f}{\partial x} \sin \left( K_0 x \sec \theta \right) \exp{ K_0 z \sec ^2 \theta } {\rm d} x \rm{d} z ,\\
	&	Q(\theta) = \iint_{S_c} \frac{\partial f}{\partial x} \cos \left( K_0 x \sec \theta \right) \exp{ K_0 z \sec ^2 \theta } {\rm d} x \rm{d} z ,
\end{align*}
where $f$ represents the ship hull form, and $f(x,z)$ represents the $y$ coordinates of the ship hull.

\subsection{Mathematical hull form dataset}\label{sec.data}
The generalized Wigley hull form is a mathematical hull form proposed by \cite{Matsui19,Matsui22} that expands the modified Wigley hull form. 
It defines a hull form using the following parameters: principal dimensions (length $L$, width $B$, and draft $d$) and coefficients of fineness (block coefficient $C_b$, midship area coefficient $C_m$, and waterplane area coefficient $C_w$). 
In the present study, we generated many hull forms using the generalized Wigley hull form by changing the parameters and subsequently using them for training the GAN model. 

The generalized Wigley hull form is defined as follows:
\begin{subequations}
	\begin{align}
		\eta &= f\left( \xi, \zeta \right), \\
		&= \left( 1- \zeta^{Z_1} \right) \left( 1- \xi^{X_1} \right) + \zeta^{Z_1} \left( 1-\zeta^{Z_2} \right) \left( 1- \xi^{X_2} \right)^{X_3}, \\
		{\rm where} \notag \\
		C_p &= \frac{C_b}{C_m} ,\\
		X_1 &= \frac{C_w}{1-C_w}, \\
		X_2 &= \max \left( 2, \frac{C_p}{1-C_p} \right), \\
		X_3	&= \frac{1}{ {C_p} ^2}, \\
		Z_1 &= \frac{C_b - SC_m}{ C_w - C_b - S\left( 1-C_m \right) }, \\
		Z_2 &= \frac{C_m}{1-C_m}\frac{C_w - C_p}{C_w - C_b - S \left( 1-C_m \right)}, \\
		S 	&= \frac{ \gamma \left( \frac{1}{X_2} + 1 \right) \gamma \left( X_3 + 1 \right) }{\gamma \left( \frac{1}{X_2} + X_3 +1 \right)}.
	\end{align} \label{eq.gw}
\end{subequations}
$\eta = f \left( \xi, \zeta \right)$ represent the $y$-coordinates of $(x,z) = (\xi, \zeta)$.
The ship hull form data are organized as vectors as follows: 
\begin{align*}
	&d = \left \{ y_{i,j}  =  f\left(x_i, z_j \right)  \mid i \in \{ 1,2,\dots,20\},~ j \in \{ 1,2,\dots,40\} \right\}, \\
	&x_i \in \left \{ 0.0, 0.2, 0.4, 0.45, 0.5, 0.55, 0.6, 0.65, 0.7, 0.75, 0.8, 0.85,  \right. \\
	& \qquad \left.  0.9, 0.925, 0.9375, 0.95, 0.9625, 0.975, 0.9875, 1.0 \right\}
\end{align*}
$x_i$ are fixed values among all data. 
Subsequently, $d$ is input into the neural network model.

\begin{figure}[h]
	\begin{center}
		\begin{minipage}[h]{\textwidth}
			\begin{center}
				\includegraphics[width=100mm]{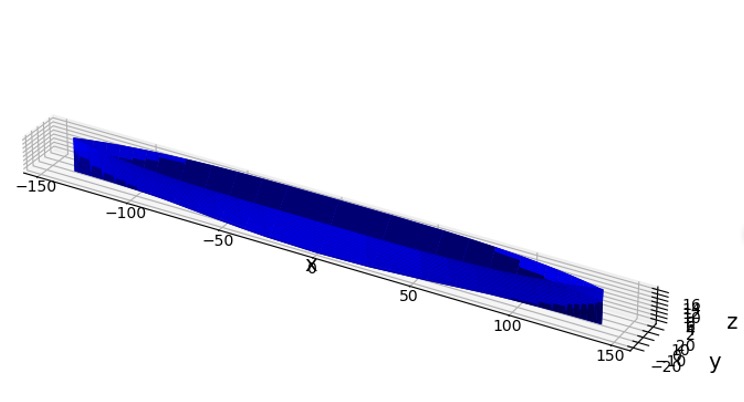}
				\par
				{(a) High-speed ship.}
			\end{center}
		\end{minipage}%
		\par
		\begin{minipage}[h]{\textwidth}
			\begin{center}
				\includegraphics[width=100mm]{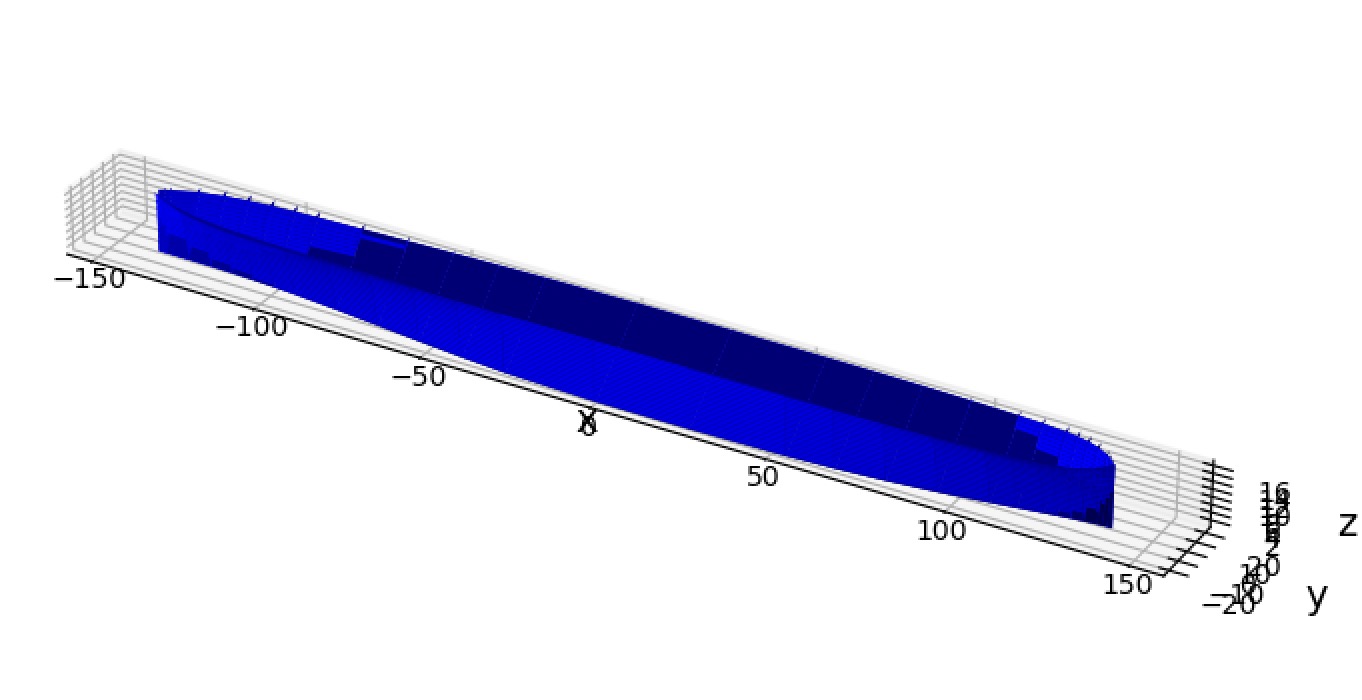}
				\par
				{(b) Mid-speed ship.}
			\end{center}
		\end{minipage}%
		\par
		\begin{minipage}[h]{\textwidth}
			\begin{center}
				\includegraphics[width=100mm]{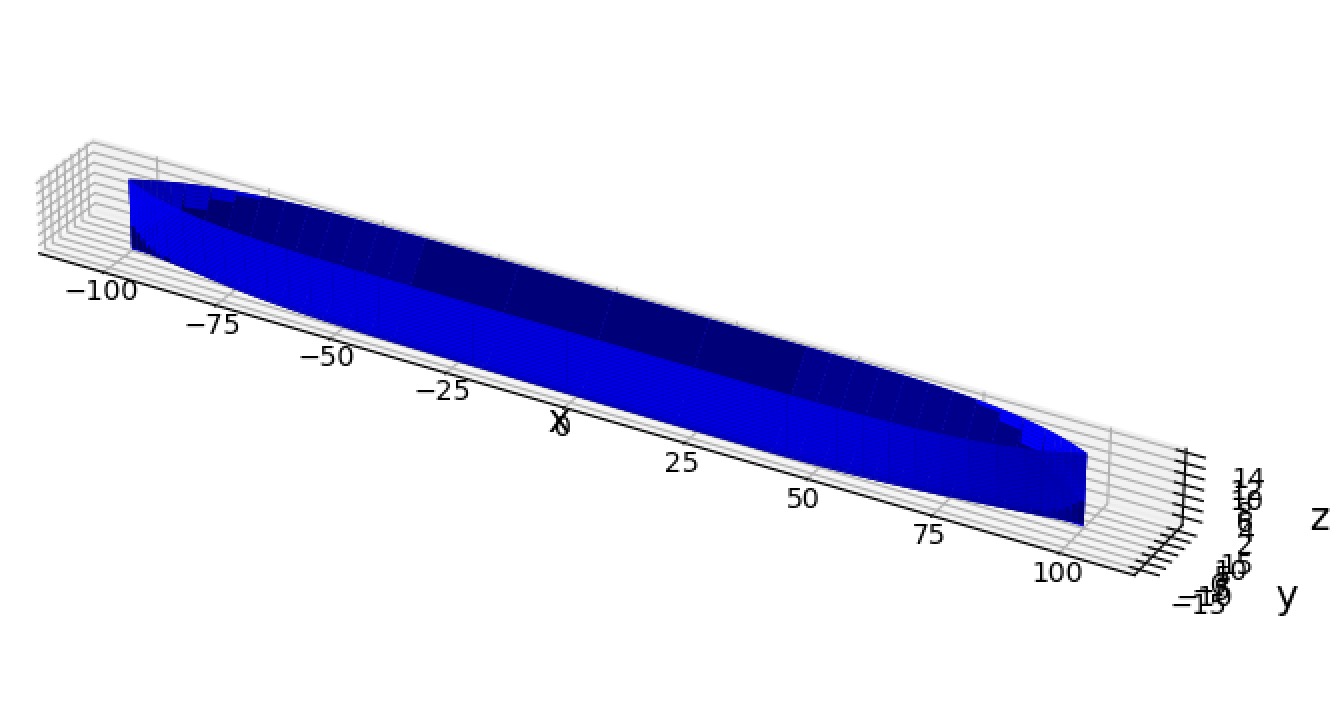}
				\par
				{(c) Low-speed ship.}
			\end{center}
		\end{minipage}%
		\caption{Examples of the line plan of the ships.}
		\label{fig:ex}
	\end{center}
\end{figure}

The training dataset was generated by selecting the parameters listed in \reftab{tab:data}, which are the typical values of high-, mid-, and low-speed ships. 
The ship–hull coordinates are generated using \refeq{eq.gw}, and $C_d$ and $W$ were calculated using \refeq{eq.cd}.
Examples of the generated shapes are presented in \reffig{fig:ex}. 

\begin{table}[h]
	\begin{center}
		\scriptsize
		\caption{Training dataset}
		\label{tab:data}
		\begin{tabular}{c|c|c|c}
			Category & 	High speed &  Middle speed & Low speed\\
			\hline
			Design speed		& 25 knot	& 20 knot 	& 15 knot \\
			B/L   	& $\{0.125\}$	&  $\{0.130, 0.135, \dots, 0.150\}$ 	& $\{0.155, 0.160, \dots, 0.200\}$ \\
			d/L 	& $\{0.045\}$	& $\{0.055, 0.060, 0.060\}$ 	& $\{0.065, 0.070\}$ \\
			$C_m$		&  $\{0.85, 0.86, \dots, 0.97\}$ & $\{0.98, 0.99\}$	& $\{0.95\}$  \\
			$C_w$		&  $\{0.50, 0.51, \dots, 0.60\}$ & $\{0.65, 0.66, \dots, 0.75\}$	& $\{0.78, 0.79, \dots, 0.85\}$  \\
			$C_b$		&  $\{0.68, 0.69, \dots, 0.78\}$ & $\{0.80, 0.81, \dots, 0.85\}$	& $\{0.86, 0.87, \dots, 0.92\}$  \\
			Number of data		& 1552	& 1594 	& 920 \\
		\end{tabular}
	\end{center}
\end{table}

\subsection{Generative model for ship hull design}
\begin{figure*}[h]
	\begin{center}
		\begin{minipage}[h]{\textwidth}
			\begin{center}
				\scalebox{0.12}{\includegraphics{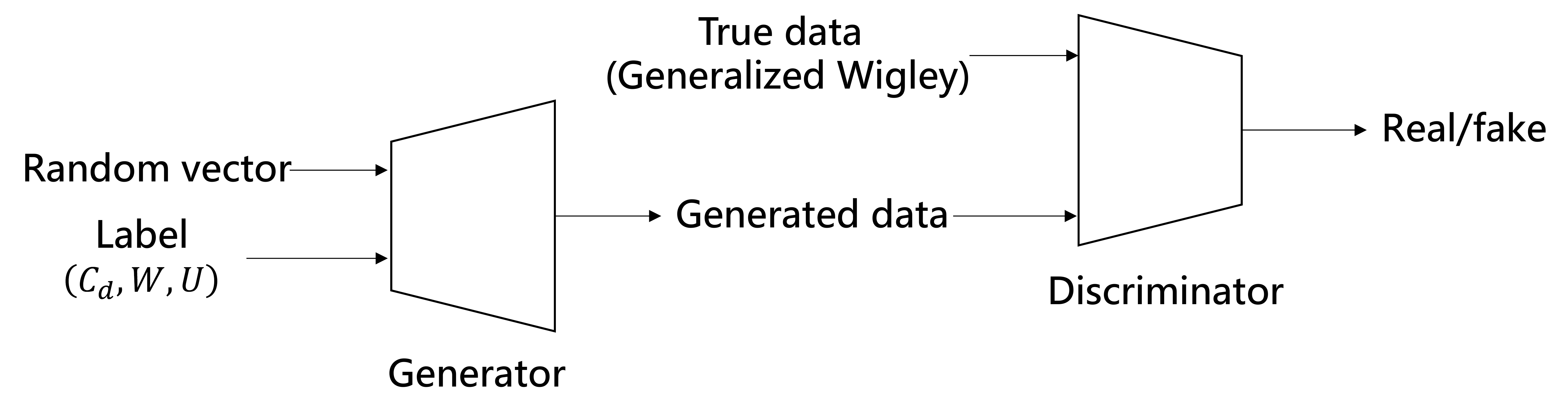}}
				\par
				{(a) GAN network.}
			\end{center}
		\end{minipage}%
		\par
		\begin{minipage}[h]{\textwidth}
			\begin{center}
				\includegraphics[height=40mm]{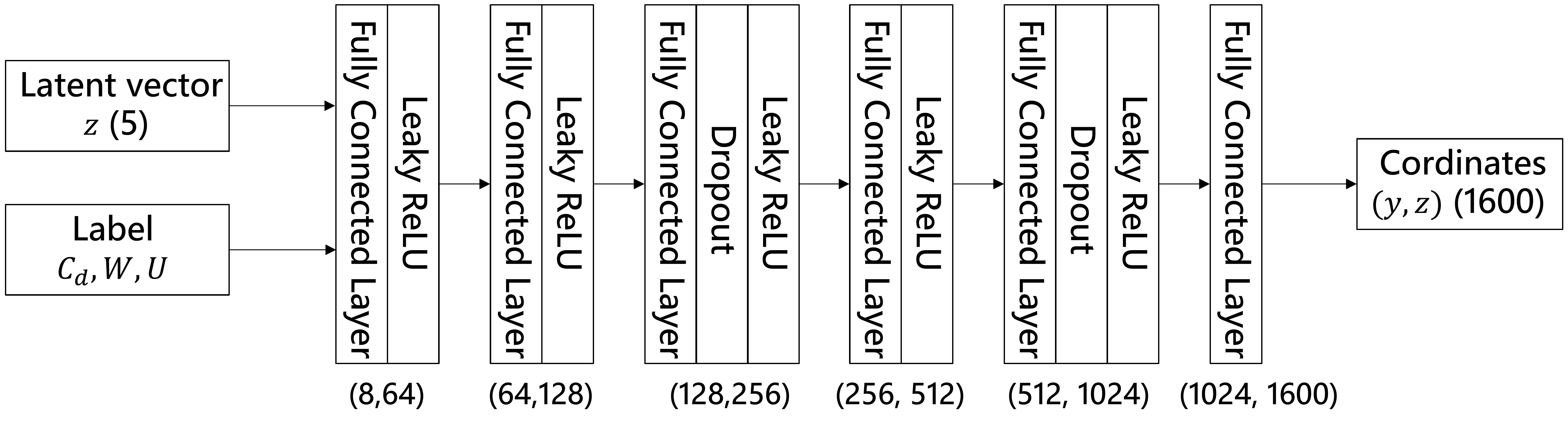}
				\par
				{(b) Generator network.}
			\end{center}
		\end{minipage}%
		\par
		\begin{minipage}[h]{\textwidth}
			\begin{center}
				\includegraphics[height=40mm]{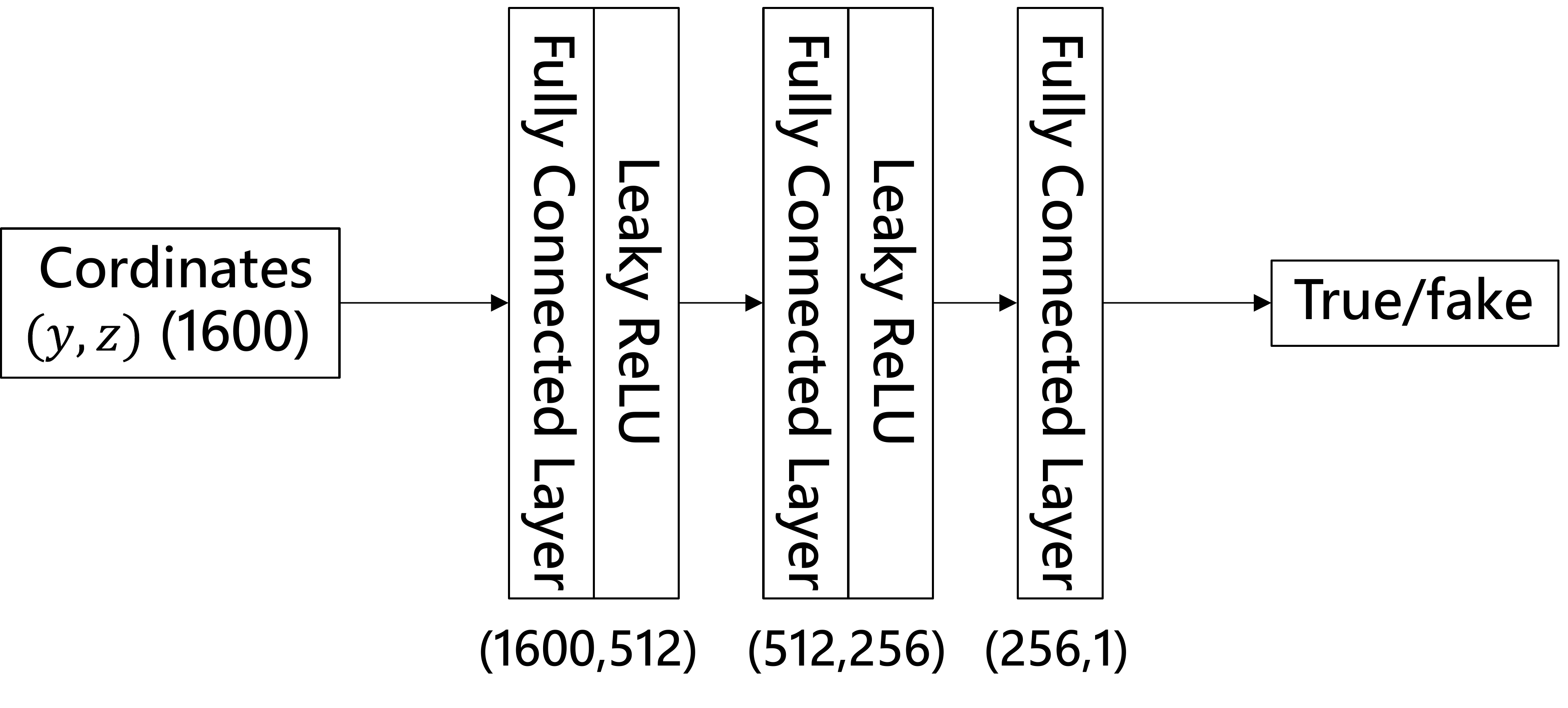}
				\par
				{(c) Discriminator network.}
			\end{center}
		\end{minipage}%
		\caption{Architecture of GAN.}
		\label{fig:arch}
	\end{center}
\end{figure*}

The architecture of the GAN model is shown in \reffig{fig:arch}. 
The generator network inputs a random noise vector $\bi{z}$ and label $ (C_d, W, U ) $, where $U$ denotes the designed cruise speed. The generator network outputs the coordinates $(y, z) \in \mathbb{R}^{1600}$ of a ship hull. 
The discriminator network inputs the coordinates $(y, z) \in \mathbb{R}^{1600}$ and outputs whether the input is true data or not. 
The number of nodes and layers are shown in \reffig{fig:arch}. 

First, the GAN model was trained using the generalized Wigley ship hull data. 
Subsequently, only the trained generator network was used to output new data. 
The random vector $\hat{\bi{z}}$ and required $(C_d, W, U)$ were input to the generator network, and the new data were output from the generator network, which outputs data that satisfy the requirements. By changing the requirement labels, the output of the generator changes to satisfy the requirements. 
In addition, the generator network outputs different data by changing the random vector, which enables the output to change while satisfying the requirements. 

The GAN model was trained based on the data and learned the correlations between the ship hull data and $(C_d, W, U)$. The GAN model did not consider physics. 
The output data were not guaranteed to meet the requirements. 
Hence, labels $(C_d, W, U)$ were recalculated using the output data and compared with the required labels.

\section{Numerical experiments}
\subsection{Experimental settings}
First, a preliminary example is presented in Section 4.1. 
A GAN model was trained without a geometric penalty function using mid-speed ship data. 
Subsequently, three GAN models with the geometric penalty function were trained using low-, middle-, and high-speed ship data, respectively. 
Finally, one GAN model was trained using all the data with the geometric penalty function. 

The generated shapes were evaluated using the reproduction of $C_d$ and $W$; $C_d$ and $W$ were recalculated from the generated data, and the mean absolute percentage error (MAPE) was calculated as follows: 
\begin{align*}
	{\rm MAPE} = \frac{1}{n} \sum_{i=1}^{n} \left |  \frac{\hat{c}_i - c_i}{c_i} \right|
\end{align*}
where $n$ is the number of samples, and $\hat{c}_i$ is the calculated value of the generated shape, where $c_i$ is the label.

The computation was conducted on an Intel Core i9 processor with 64 GB of memory equipped with an RTX 4090 GPU. 
The codes were implemented using Python \cite{python} and PyTorch \cite{PyTorch}. 

\subsection{Training cWGAN-gp using all data}\label{sec.all}

\begin{figure*}[h]
	\begin{center}
		\begin{minipage}[h]{\textwidth}
			\begin{center}
				\includegraphics[width=\textwidth]{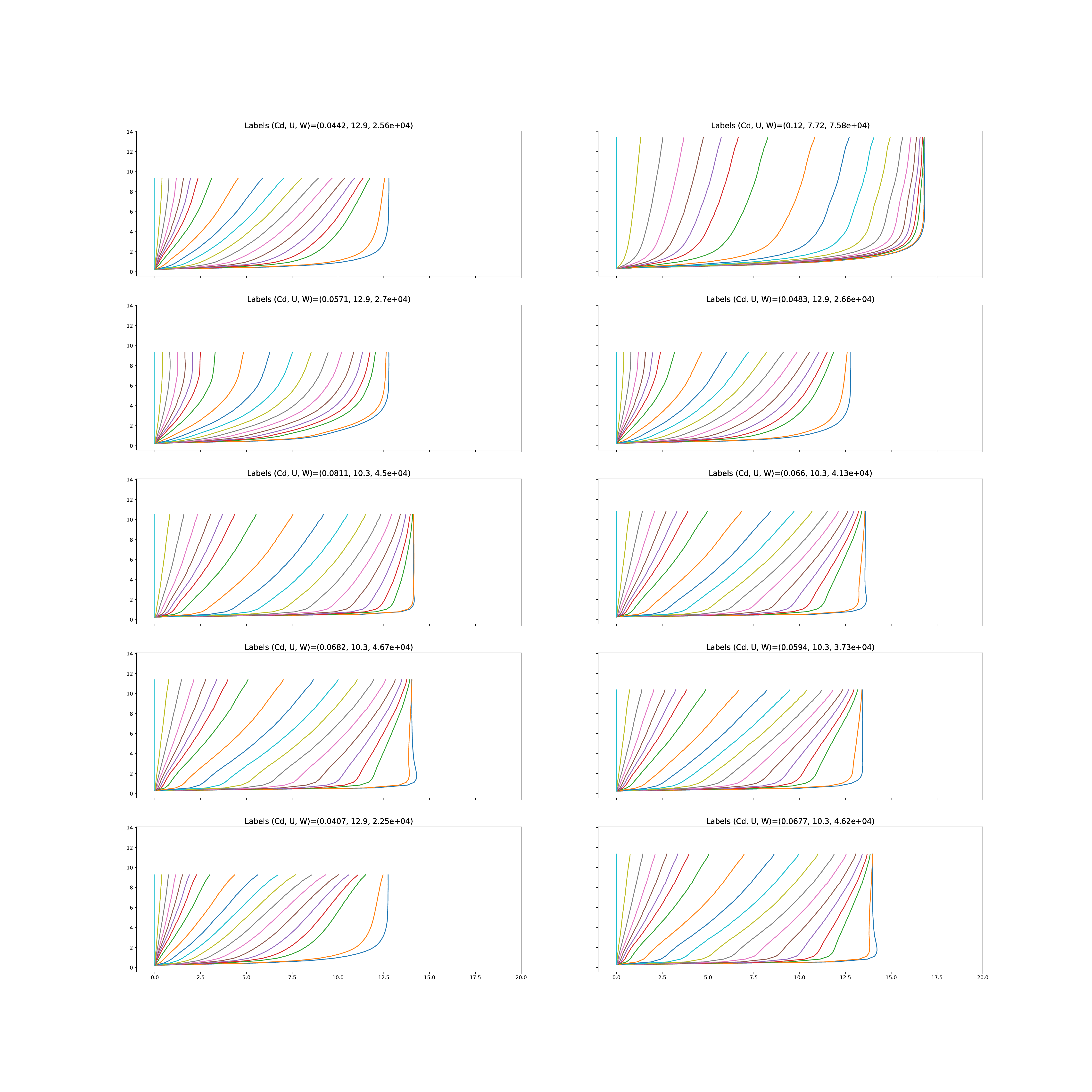}
			\end{center}
		\end{minipage}%
		\caption{Examples of the generated data for all ships of various speeds.}
		\label{fig:ex.all}
	\end{center}
\end{figure*}

\begin{figure*}[h]
	\begin{center}
		\begin{minipage}[h]{0.5\textwidth}
			\begin{center}
				\includegraphics[height=50mm]{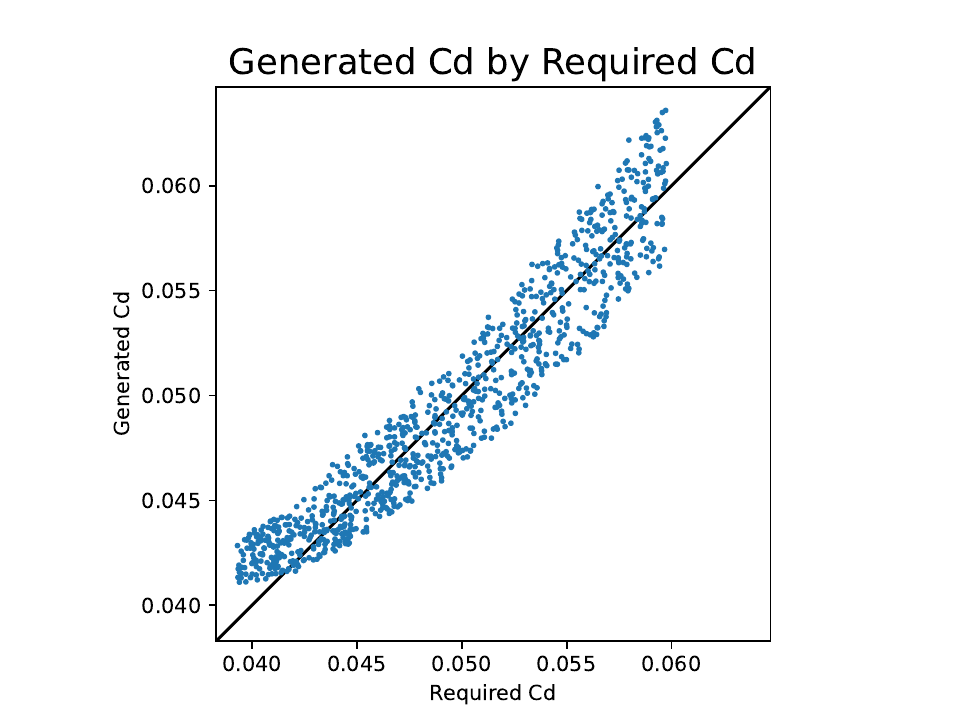}
				\par
				{(a) High-speed ship, $C_d$.}
			\end{center}
		\end{minipage}%
		\begin{minipage}[h]{0.5\textwidth}
			\begin{center}
				\includegraphics[height=50mm]{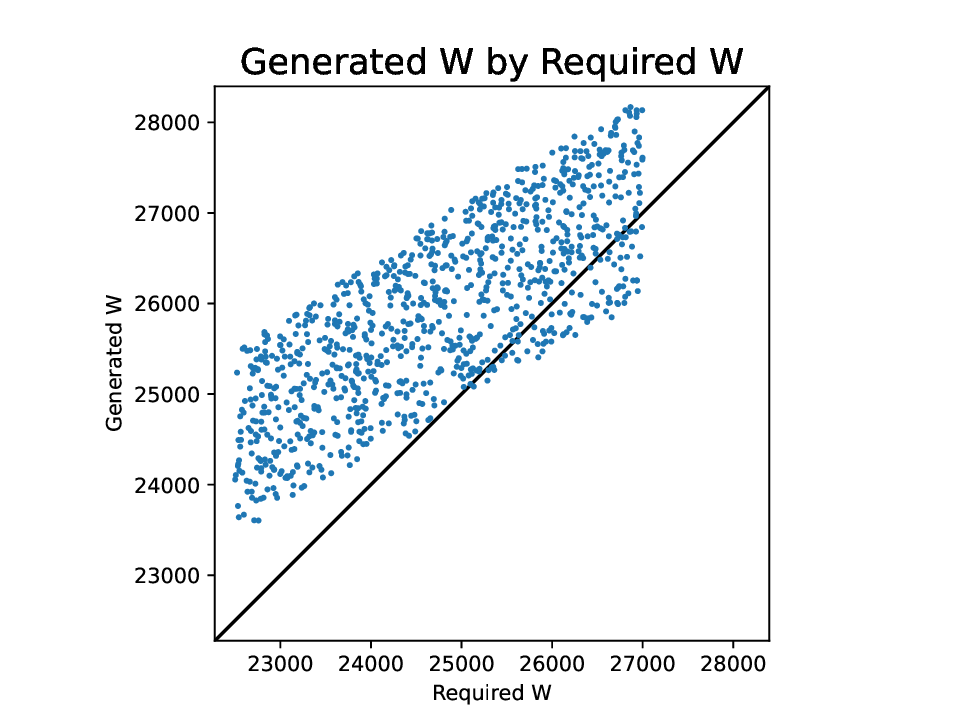}
				\par
				{(b) High-speed ship, $W$.}
			\end{center}
		\end{minipage}%
		\par
		\begin{minipage}[h]{0.5\textwidth}
			\begin{center}
				\includegraphics[height=50mm]{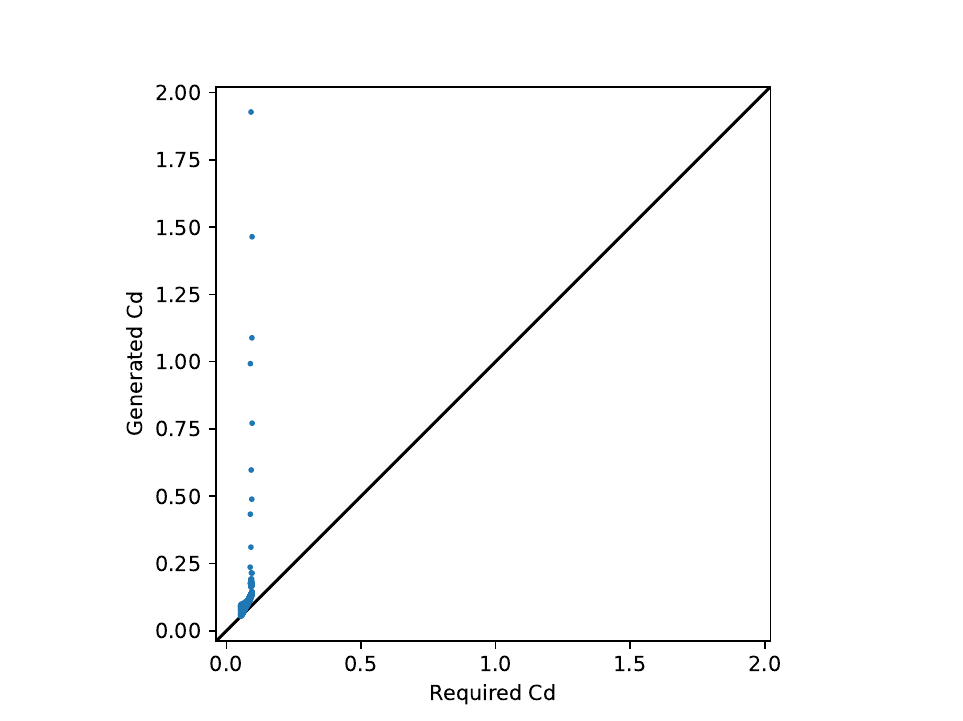}
				\par
				{(c) Mid-speed ship, $C_d$.}
			\end{center}
		\end{minipage}%
		\begin{minipage}[h]{0.5\textwidth}
			\begin{center}
				\includegraphics[height=50mm]{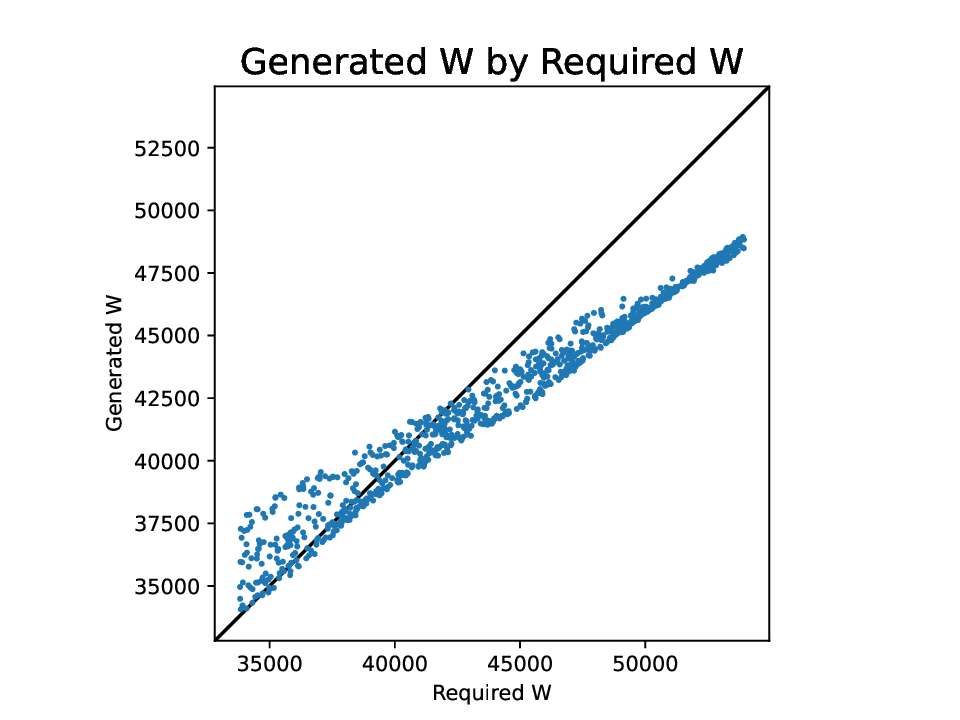}
				\par
				{(d) Mid-speed ship, $W$.}
			\end{center}
		\end{minipage}%
		\par
		\begin{minipage}[h]{0.5\textwidth}
			\begin{center}
				\includegraphics[height=50mm]{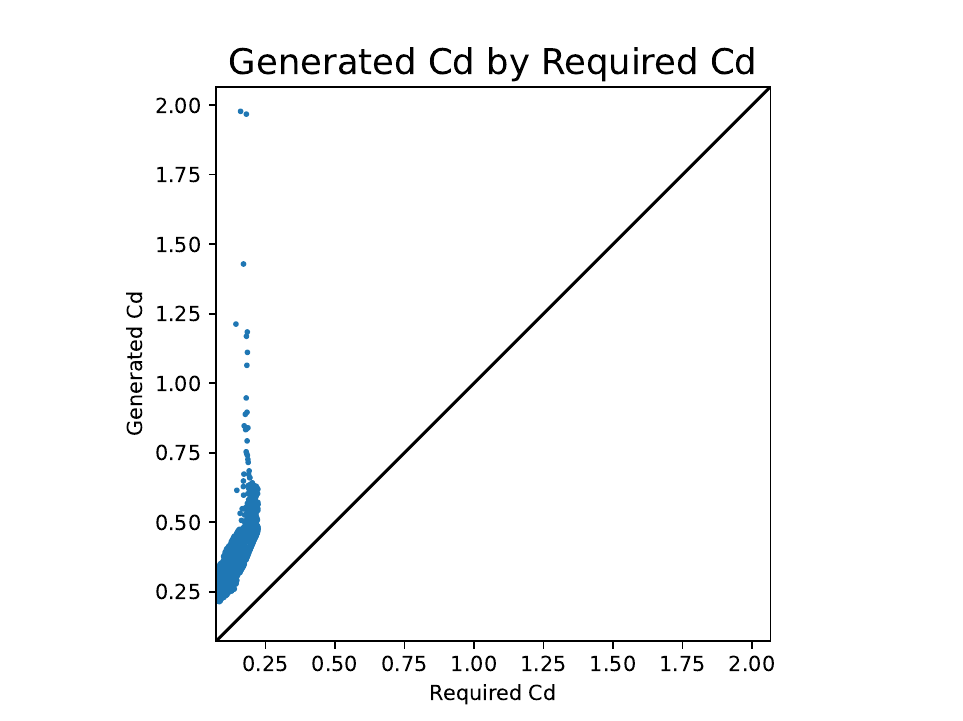}
				\par
				{(e) Low-speed ship, $C_d$.}
			\end{center}
		\end{minipage}%
		\begin{minipage}[h]{0.5\textwidth}
			\begin{center}
				\includegraphics[height=50mm]{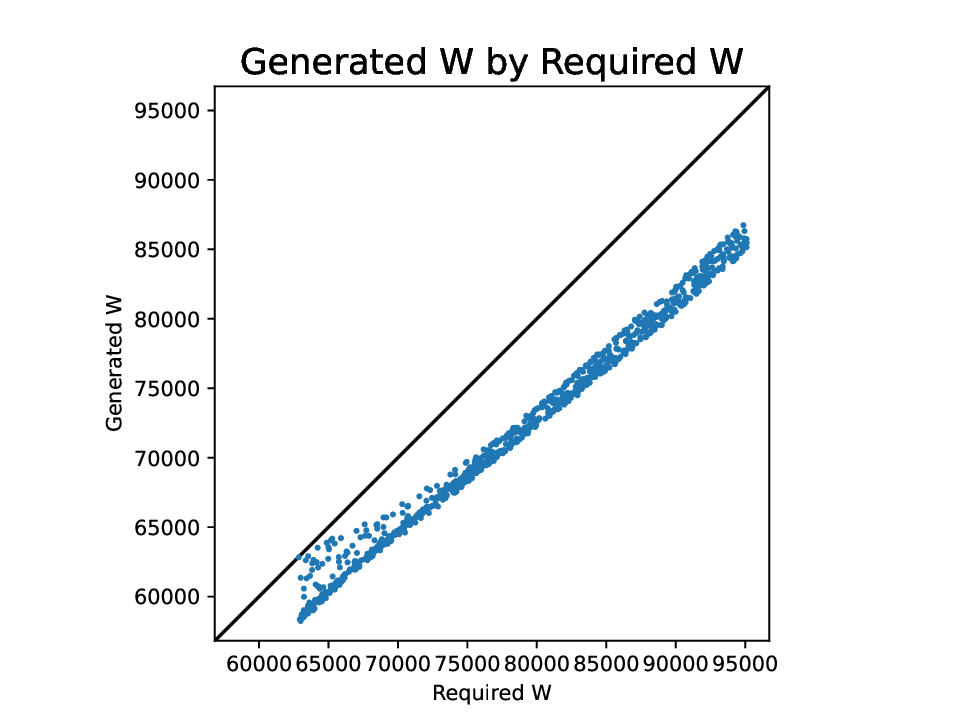}
				\par
				{(f) Low-speed ship, $W$.}
			\end{center}
		\end{minipage}%
		\caption{Scatter plots of $C_d$ and $W$ in the integrated model.}
		\label{fig:scatter_all}
	\end{center}
\end{figure*}

The CWGAN-gp model was trained using all fast-, mid-, and low-speed ship data. 
Subsequently, ship hull shapes were generated by specifying different cruise speeds, i.e., fast, middle, and low speeds, and different requirements $C_d$ and $W$. 
The output shapes are shown in \reffig{fig:ex.all}. At first glance, the output shapes appear smooth and reasonable. 
However, the MAPEs were not as low as desired. The results of MAPE are shown in \reftab{tab:mape_all}, and the scatter plots of the labels and recalculated labels are shown in \reffig{fig:scatter_all}. 
The MAPEs were large, particularly for middle- and low-speed ships. 
The scatter plot also indicates that the error is large for the middle- and low-speed ships. 
The reason for the large MAPEs is that the model was trained using all data that contained different data styles, i.e., fast-, middle-, and low-speed ships. 
Processing and distinguishing between these different data was difficult. 

\begin{table}
	\begin{center}
		\caption{MAPE of the generated ship hull form of the integrated model.}
		\label{tab:mape_all}
		\begin{tabular}{l|r|r|r}
			Design speed & 	MAPE of $C_d$	& MAPE of $W$ & Total \\
			\hline
			High speed   	& 0.03171	& 0.05069	& 0.04120 \\
			Medium speed 	& 0.38770	& 0.05002	& 0.21886 \\
			Low speed   	& 1.74973	& 0.08326 	& 0.91650 \\
		\end{tabular}
	\end{center}
\end{table}

\subsection{Training cWGAN-gp using separate data}
\begin{figure*}[h]
	\begin{center}
		\begin{minipage}[h]{0.5\textwidth}
			\begin{center}
				\includegraphics[height=50mm]{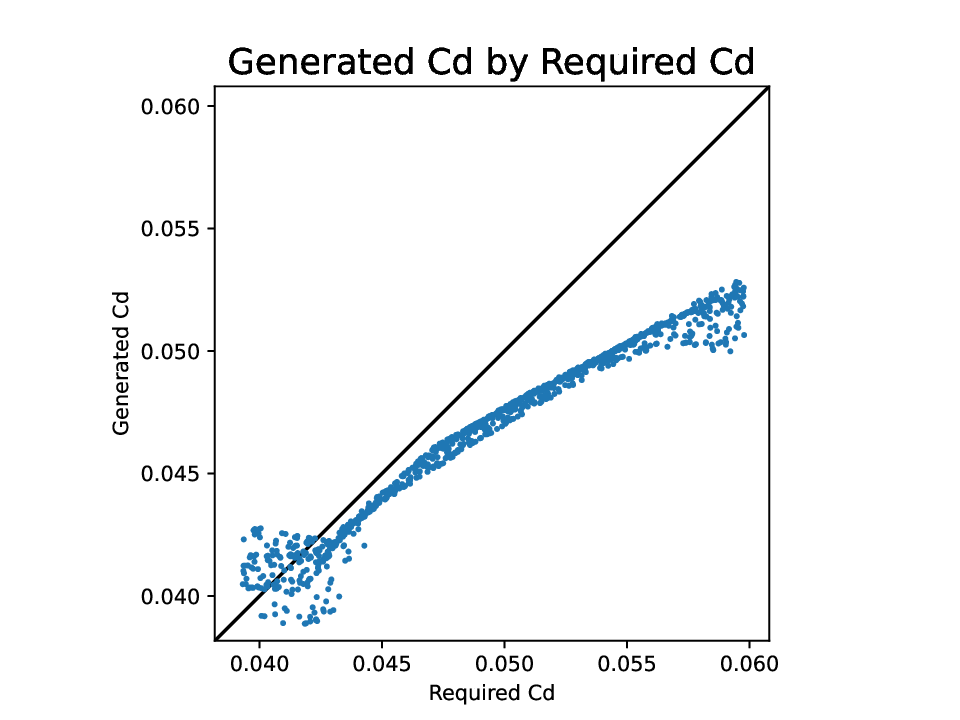}
				\par
				{(a) High-speed ship, $C_d$.}
			\end{center}
		\end{minipage}%
		\begin{minipage}[h]{0.5\textwidth}
			\begin{center}
				\includegraphics[height=50mm]{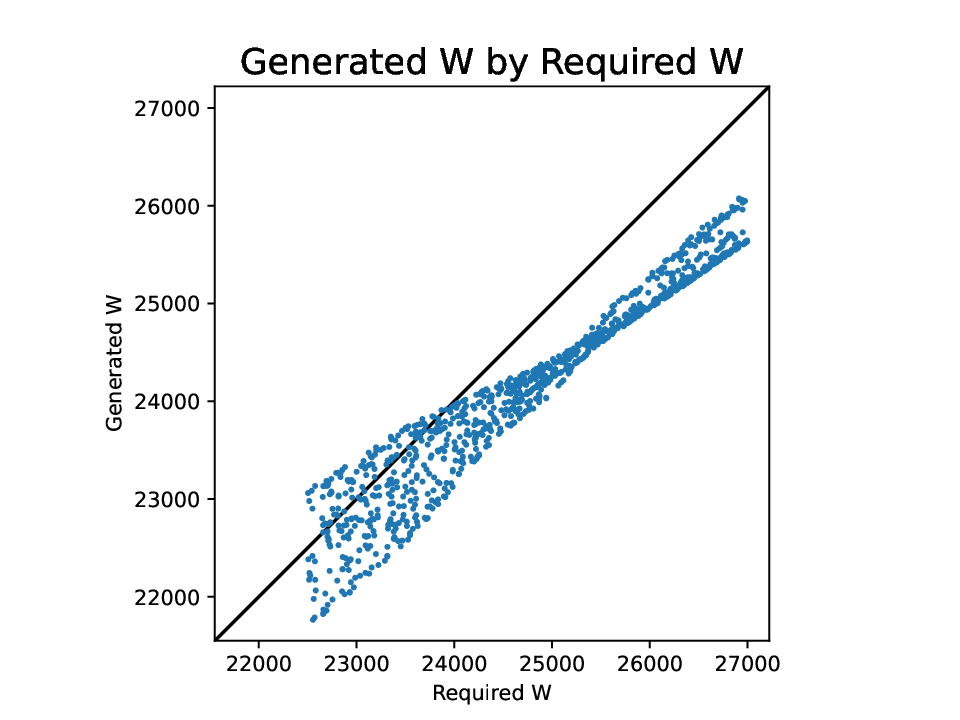}
				\par
				{(b) High-speed ship, $W$.}
			\end{center}
		\end{minipage}%
		\par
		\begin{minipage}[h]{0.5\textwidth}
			\begin{center}
				\includegraphics[height=50mm]{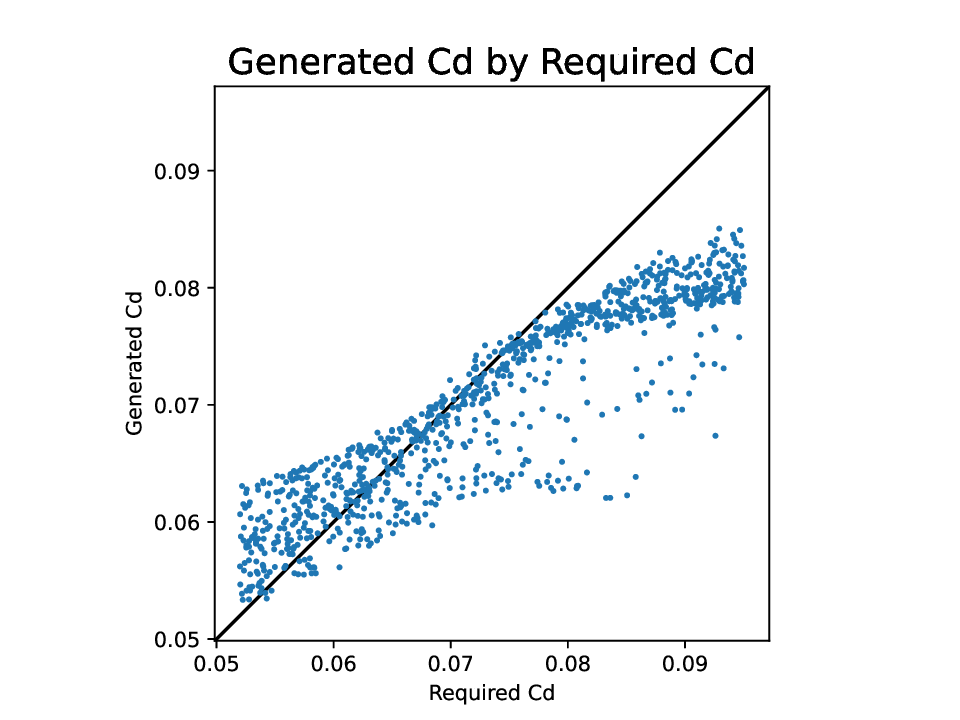}
				\par
				{(c) Mid-speed ship, $C_d$.}
			\end{center}
		\end{minipage}%
		\begin{minipage}[h]{0.5\textwidth}
			\begin{center}
				\includegraphics[height=50mm]{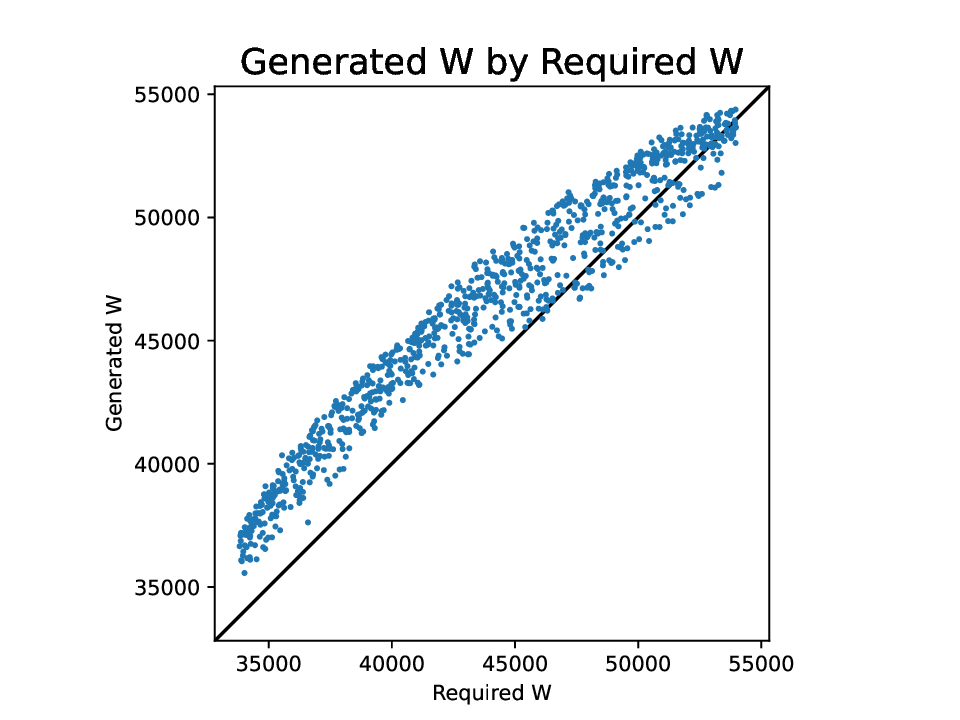}
				\par
				{(d) Mid-speed ship, $W$.}
			\end{center}
		\end{minipage}%
		\par
		\begin{minipage}[h]{0.5\textwidth}
			\begin{center}
				\includegraphics[height=50mm]{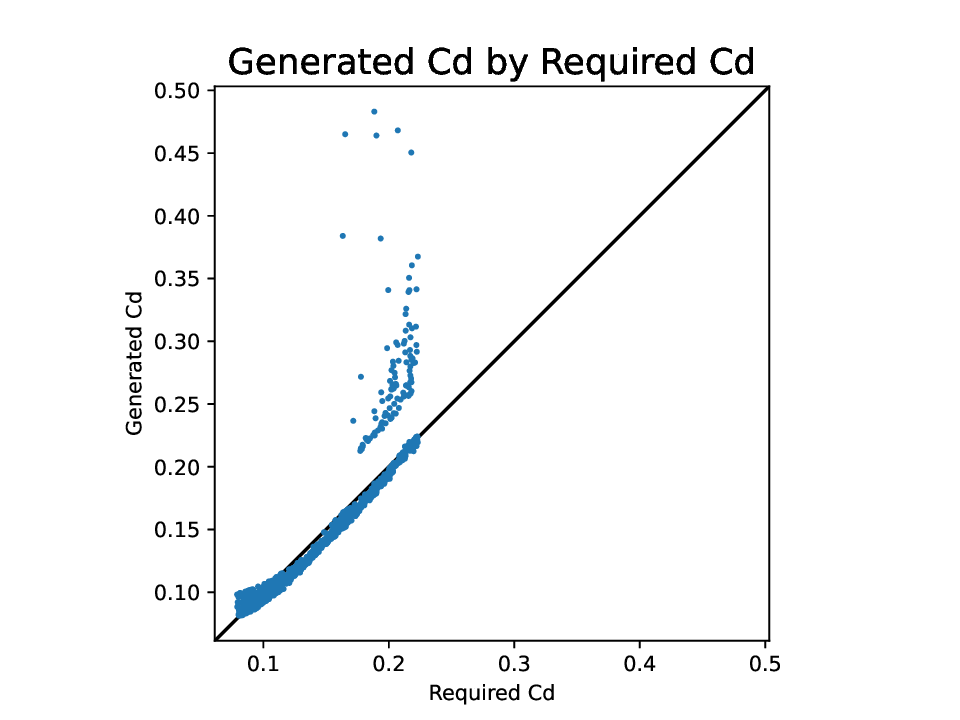}
				\par
				{(e) Low-speed ship, $C_d$.}
			\end{center}
		\end{minipage}%
		\begin{minipage}[h]{0.5\textwidth}
			\begin{center}
				\includegraphics[height=50mm]{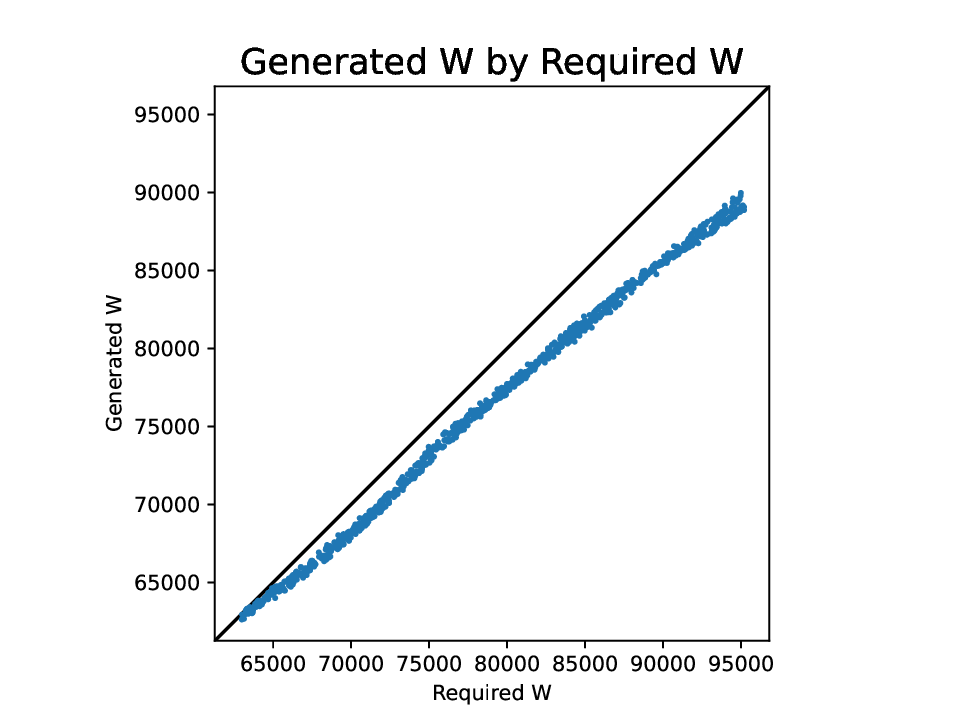}
				\par
				{(f) Low-speed ship, $W$.}
			\end{center}
		\end{minipage}%
		\caption{Scatter plots of $C_d$ and $W$.}
		\label{fig:scatter}
	\end{center}
\end{figure*}

The data were separated into fast-, middle-, and low-speed data and separately fed into three GAN models; each GAN model was trained with one-speed ship data. 	
The MAPEs of $C_d$ and $W$ are listed in \reftab{tab:mape_sep} and \reffig{fig:scatter}, respectively. 
The MAPEs differ for different cruise speeds and are less than 0.09. 
The MAPEs are smaller than those of the integrated model explained in Section \ref{sec.all}, and by using separate data, the MAPE was decreased. 

The numerical results also show that the MAPE of $W$ is lower than that of $C_d$. This is because $W$ is directly related to the geometry, whereas the relationship between $C_d$ and ship hull geometry is complicated, as formulated in \refeq{eq.cd}.  

\begin{figure*}[h]
	\begin{center}
		\begin{minipage}[h]{\textwidth}
			\begin{center}
				\includegraphics[width=\textwidth]{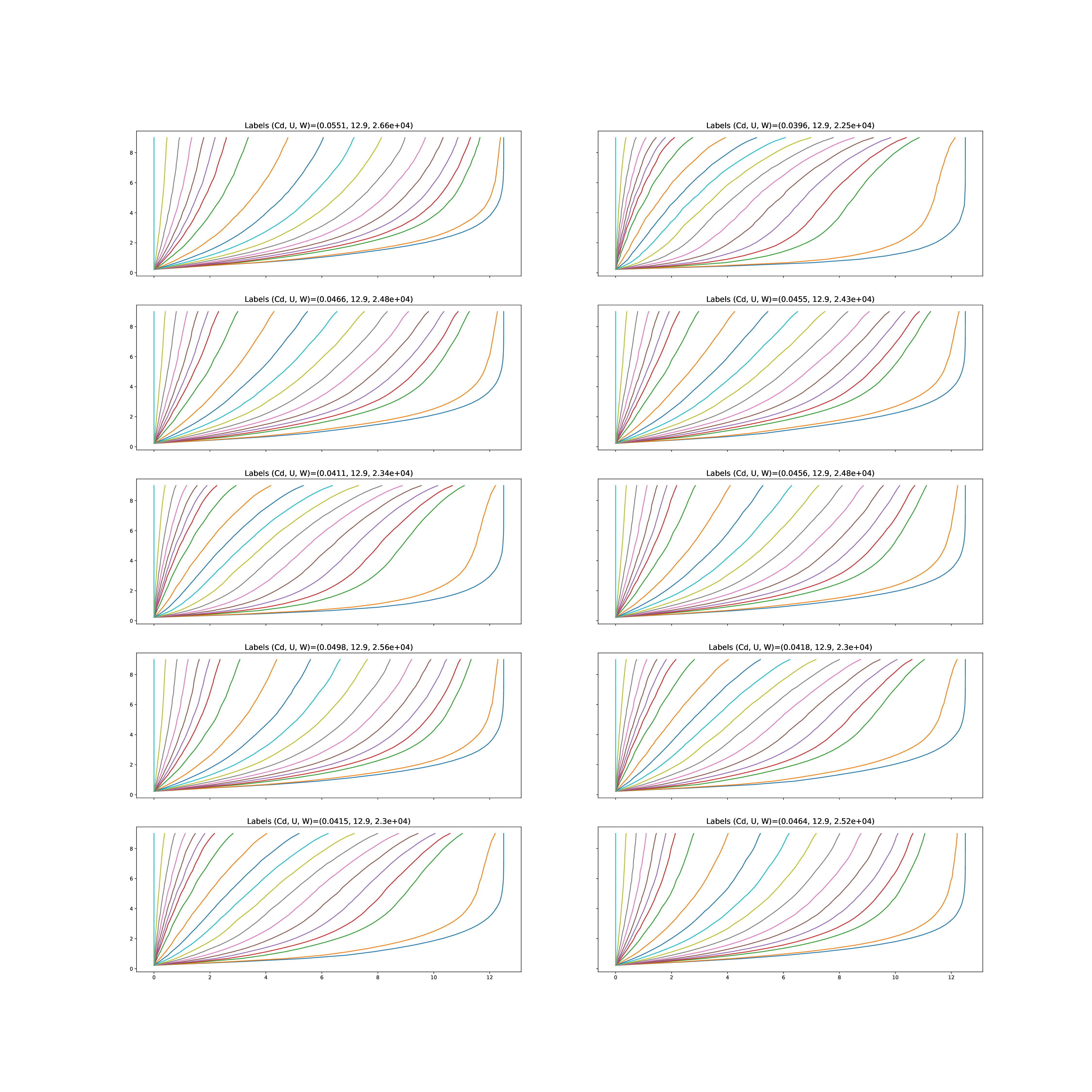}
			\end{center}
		\end{minipage}%
		\caption{Examples of the generated data for high-speed ships.}
		\label{fig:ex.high}
	\end{center}
\end{figure*}

\begin{figure*}[h]
	\begin{center}
		\begin{minipage}[h]{\textwidth}
			\begin{center}
				\includegraphics[width=\textwidth]{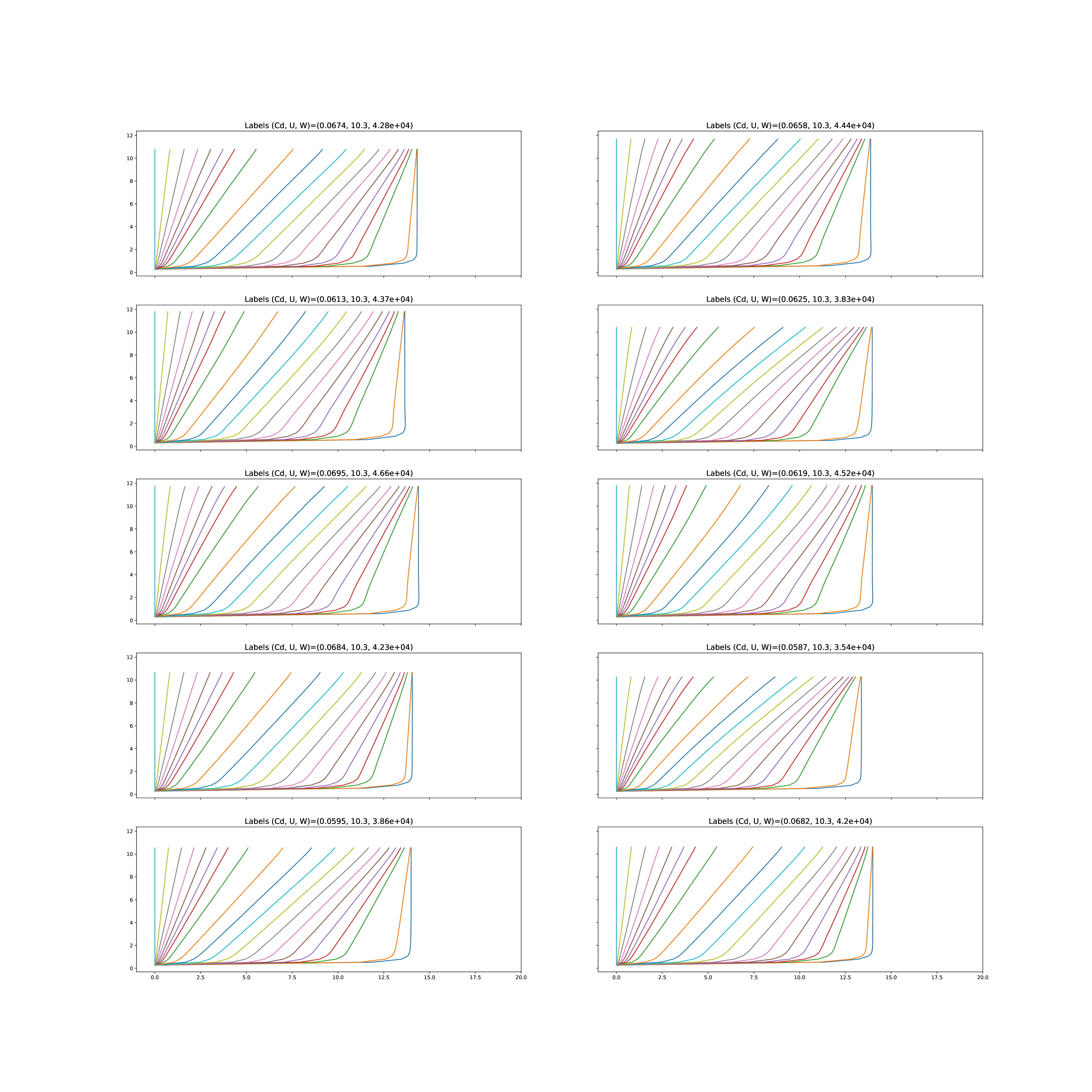}
			\end{center}
		\end{minipage}%
		\caption{Examples of the generated data for mid-speed ships.}
		\label{fig:ex.mid}
	\end{center}
\end{figure*}

\begin{figure*}[h]
	\begin{center}
		\begin{minipage}[h]{\textwidth}
			\begin{center}
				\includegraphics[width=\textwidth]{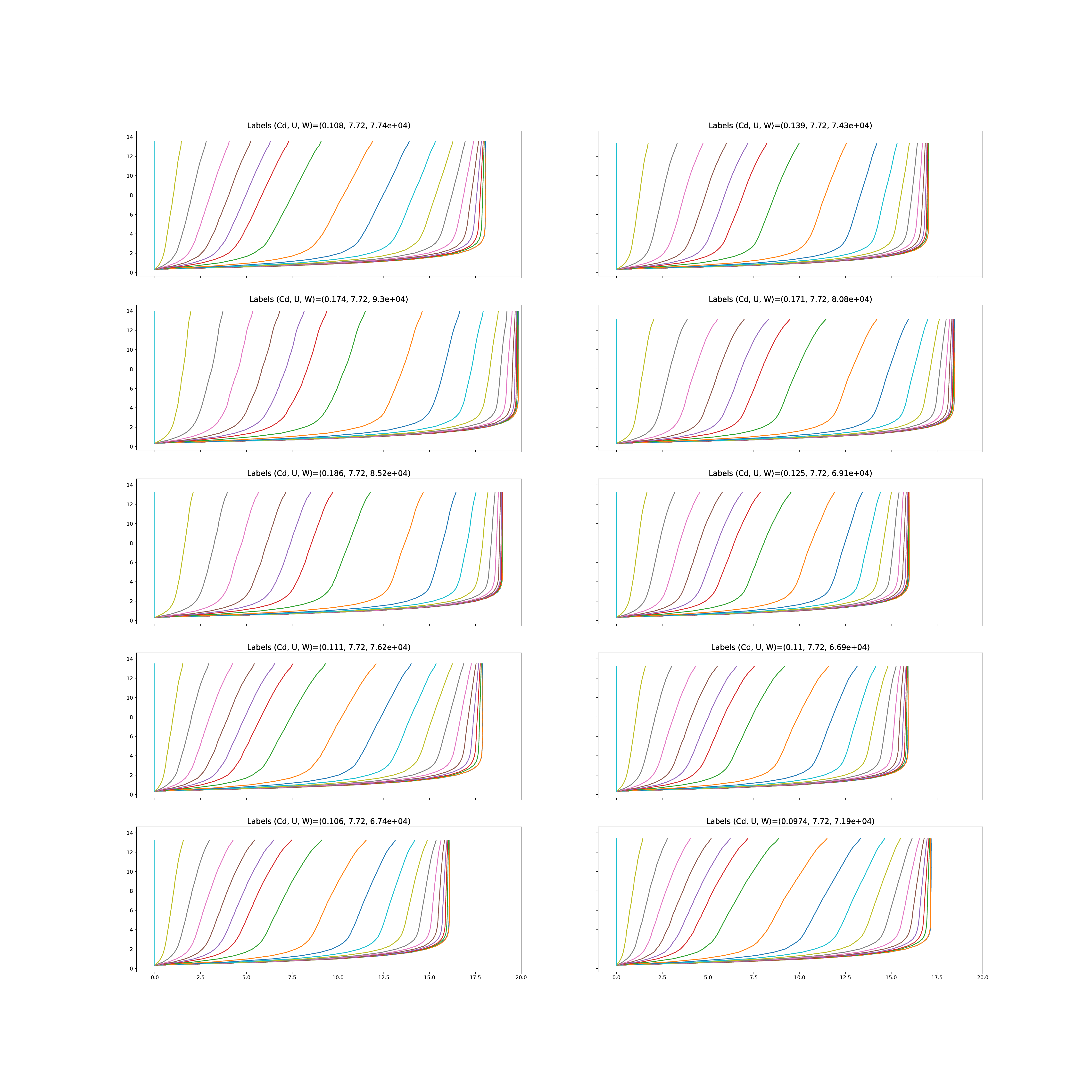}
			\end{center}
		\end{minipage}%
		\caption{Examples of the generated data for low-speed ships.}
		\label{fig:ex.low}
	\end{center}
\end{figure*}

Examples of the generated ship hull data are presented in \reffig{fig:ex.high}, \reffig{fig:ex.mid}, and \reffig{fig:ex.low}. 
The ship hull shapes exhibit smooth curves. The generated ships were wider for low-speed ships and thinner for high-speed ships. A wider width leads to large $C_d$ and large $W$, which is consistent with the characteristics of low-speed ships.
In addition, low-speed ships have a square shape, whereas high-speed ones have more round-like shapes, which are also related to $Cd$ and $W$. 
The output data exhibit different shapes. For example, high-speed ships are thinner than low-speed ships, which is reasonable.

\begin{table}
	\begin{center}
		\caption{MAPEs of the generated ship hull form.}
		\label{tab:mape_sep}
		\begin{tabular}{l|r|r|r}
			Design speed & 	MAPE of $C_d$	& MAPE of $W$ & Total \\
			\hline
			High speed   	& 0.04347	& 0.07327 	& 0.05837 \\
			Medium speed 	& 0.07061	& 0.06144	& 0.06603 \\
			Low speed   	& 0.08452	& 0.03362 	& 0.05907 \\
		\end{tabular}
	\end{center}
\end{table}

\section{Conclusion}
A ship hull design method using cWGAN-gp was proposed. The proposed model was used to generate ship hull data by specifying the ship performance, such as the drag coefficient, which is a more direct requirement than geometrical requirements, such as block coefficient $C_b$ and midship area coefficient $C_m$. 
The proposed model successfully generated new ship hull data. 
Fast-, middle-, and low-speed data were prepared and fed into the GAN model.
Using all the data at once to train the model led to a large MAPE. 
However, the MAPE was decreased by separating the data and training different GAN models. 
This result implies that the ship hull data for different design cruise speeds have different features, and processing different data using different GAN models is easier than using all the data at once. 
The numerical results also showed that the MAPE of $W$ is smaller than that of $C_d$, implying that the labels directly related to the geometry are easier to process than those with a more complicated relationship with ship hull geometry. 

\section*{Acknowledgement}
This work was supported by JSPS KAKENHI Grant Numbers JP21K14064 and 23K13239.







\end{document}